\documentstyle[12pt]{article}
\begin{document}

\def\frak{\cal }
\def\Bbb{\bf }

\title{Quantum interfaces}
\author{Karl Svozil\\
 {\small Institut f\"ur Theoretische Physik,}
  {\small Technische Universit\"at Wien }     \\
  {\small Wiedner Hauptstra\ss e 8-10/136,}
  {\small A-1040 Vienna, Austria   }            \\
  {\small e-mail: svozil@tuwien.ac.at}}
\date{ }
\maketitle

\begin{flushright}
{\scriptsize
http://tph.tuwien.ac.at/$\widetilde{\;\;}\,$svozil/publ/interface.$\{$htm,ps,tex$\}$}
\end{flushright}

\begin{abstract}
Rational agents acting as observers use ``knowables''
to construct a vision of the outside
world.
Thereby, they are bound by the information exchanged
with what they consider to be objects.
The cartesian cut or, in modern terminology, the interface mediating
this exchange, is again a construction.
It serves as a ``scaffolding,''
an intermediate construction capable of providing the necessary conceptual means.

An attempt is made to formalize the interface, in particular the quantum interface
and quantum measurements, by
a symbolic information exchange.
A principle of conservation of information is reviewed and a measure
of information flux through the interface is proposed.

We cope with the question of why
observers usually experience irreversibility in measurement processes if the evolution
is reversible, i.e.,  one-to-one.
And why should there be any meaningful concept of classical
information if there is merely quantum information to begin with?
We take the position here that the concept of irreversible measurement
is no deep principle but originates in the practical inability
to reconstruct a quantum state of the object.

Many issues raised apply also to the quantum's natural double, virtual reality.

An experiment is proposed to test the conjecture that the self is transcendent.

\end{abstract}

\section*{Reality construction by ``knowables''}

Otto R\"ossler, in a thoughtful book \cite{roessler-98},
has pointed to the significance of object-observer interfaces, a topic
which had also been investigated in other contexts
(cf., among others, refs.
\cite{bos,toffoli:79,svo5,svo-86,roessler-87,svozil-93,atman:93}).
By taking up this theme, the
following investigation is on the epistemology of interfaces, in
particular of quantum interfaces.
The informal
notions of ``cartesian cut'' and ``interface'' are formalized.
They are then applied to observations of quantum and virtual reality systems.

A generic interface is presented here as any means of
communication or information
exchange between some ``observer'' and some observed ``object.''
The ``observer'' as well as the ``object'' are subsystems
of some larger, all-encompassing system called ``universe.''

Generic interfaces are totally symmetric.
There is no
principal, {\it a priori} reason to  call one subsystem ``observer'' and
the other subsystem ``object.'' The denomination is arbitrary.
Consequently,  ``observer'' and ``object'' may switch identities.

Take, for example, an impenetrable curtain  separating two
parts of the same room.
Two parties --- call them Alice and John --- are merely allowed
to communicate by sliding through papers below the curtain. Alice,
receiving the
memos emanating from John's side of the curtain, thereby effectively
constructs a ``picture'' or representation of John and {\it vice versa}.

The cartesian cut spoils this total symmetry and arbitrariness.
It defines a distinction between ``observer'' and ``object'' beyond
doubt. In our example, one agent --- say Alice --- becomes the observer
while the other agent becomes the observed object. That, however, may be
a very arbitrary convention which not necessarily reflects the
configuration properly.

A cartesian cut may presuppose a certain sense of ``rationality,'' or even
``consciousness'' on the ``observer's'' side.
We shall assume that some observer or agent exists which, endowed with
rational intelligence,
draws conclusions
on the basis of certain premises, in particular the agent's {\em state of knowledge}, or
``knowables'' to (re)construct ``reality.''
Thereby, we may imagine the agent as some kind of robot, some mechanistic
or algorithmic entity.
(From now on, ``observer'' and ``agent'' will be used as synonyms.)
Note that the agent's state of knowledge may not necessarily coincide with
a complete description of the observed system, nor may the agent be in the possession of
a complete description of its own side of the cut.
Indeed, it is not unreasonable to speculate that certain things, although knowable ``from the outside''
of the observer-object system, are principally unknowable to an intrinsic observer \cite{svozil-93}.

Although we shall come back to this issue later, the  notion
of ``consciousness''
will not be reviewed here. We shall neither speculate
exactly what  ``consciousness'' is,  nor what may be the necessary and
sufficient conditions for an agent to be ascribed ``consciousness''.
Let it suffice to refer to two proposed tests of consciousness by Turing
and Greenberger \cite{greenberger:pr}.

With regards to the type of symbols exchanged, we shall
differentiate between two classes: classical symbols, and quantum symbols.
The cartesian cuts mediating classical and quantum symbols will be called
``classical'' or ``quantum'' (cartesian) cuts, respectively.

\section*{Formalization of the cartesian cut}

The task of formalizing the heuristic notions of ``interface'' and
``cartesian cut''
 is, at least to some extent,
analogous to the formalization of the informal notion of ``computation''
and ``algorithm'' by recursive function theory via the Church-Turing thesis.

In what follows, the informal notions of interface and  cartesian cut will
be formalized by symbolic exchange; i.e., by the mutual communication of
symbols of a formal alphabet.
In this model, an object and an observer alphabet will be associated
with the observed object and with the observer, respectively.

Let there be an {\em object
 alphabet} ${\cal S}$ with  symbols $s \in {\cal S}$
associated with the outcomes or ``message'' of an experiment
possible results.
 Let there be
 an {\em observer alphabet} ${\cal T}$ with symbols $t \in {\cal T}$ associated
with the possible inputs or ``questions'' an observer can ask.

At this point we would like to keep the observer and object alphabets as general as possible,
allowing also for quantum bits to be transferred. Such quantum bits, however, have no
direct operational meaning, since they cannot be completely specified. Only classical bits
have a (at least in principle) unambiguous meaning, since they can be completely specified, copied and measured.
We shall define an interface  next.

\begin{itemize}
\item
An interface $I$ is an entity forming the common boundary between two
parts
of a system, as  well as a means of information exchange between those
parts.

\item
By convention, one part of the of the system is called ``observer''
and the other part ``object.''

\item
Information between the observer and the object via the interface is exchanged by symbols.
The corresponding functional representation of the interface is a map
$I: {\cal T} \mapsto {\cal S}$,
where  ${\cal T}$ and ${\cal S}$ are the observer and the object
alphabets, respectively.
Any such information exchange is called {\em ``measurement.''}

\item
The interface ist {\em total} in the sense that the observer receives {\em all} symbols emanating from the object.
(However, the object needs not receive all symbols emanating
from the observer.)

\item
Types of interface include purely classical, quasi-classical, and purely quantum interfaces.
\begin{itemize}

\item
Classical scenario I:
A classical interface is an interface defined in a classical system,
for which the symbols  in ${\cal S}$ and ${\cal T}$ are
classical states encodable by classical bits ``$0$'' and ``1''
corresponding to ``${\tt true}$'' and ``${\tt false}$,'' respectively.
This kind of binary code alphabet corresponds
to yes-no outcomes
to dichotomic  questions;
experimental physics in-a-nutshell.
An example for a dichotomic outcome associated with
is
``there is a click in a counter'' or ``there is no click in a counter,'' respectively.

\item
Quasi-classical scenario II:
a quasi-classical interface is an interface defined in a quantum system,
whereby the symbols in ${\cal S}$ and ${\cal T}$ are
classical states encoded by classical bits.
This is the picture most commonly used for measurements in quantum mechanics.

\item
Quantum scenario III:
A quantum interface is an interface defined in a quantized system.
In general, the quantum symbols  in ${\cal S}$ and ${\cal T}$ are
quantum states.
\end{itemize}
\end{itemize}

\begin{figure}
\begin{center}
\unitlength 1.00mm
\linethickness{0.4pt}
\begin{picture}(55.00,60.00)
\multiput(40.00,0.00)(0.12,0.31){8}{\line(0,1){0.31}}
\multiput(40.93,2.46)(0.11,0.35){7}{\line(0,1){0.35}}
\multiput(41.71,4.89)(0.10,0.40){6}{\line(0,1){0.40}}
\multiput(42.34,7.29)(0.12,0.59){4}{\line(0,1){0.59}}
\multiput(42.82,9.65)(0.11,0.78){3}{\line(0,1){0.78}}
\multiput(43.14,11.99)(0.08,1.15){2}{\line(0,1){1.15}}
\put(43.31,14.30){\line(0,1){2.28}}
\multiput(43.32,16.57)(-0.07,1.12){2}{\line(0,1){1.12}}
\multiput(43.18,18.82)(-0.10,0.74){3}{\line(0,1){0.74}}
\multiput(42.89,21.03)(-0.11,0.55){4}{\line(0,1){0.55}}
\multiput(42.45,23.22)(-0.12,0.43){5}{\line(0,1){0.43}}
\multiput(41.85,25.37)(-0.11,0.30){7}{\line(0,1){0.30}}
\multiput(41.10,27.49)(-0.11,0.25){10}{\line(0,1){0.25}}
\multiput(40.00,30.00)(-0.12,0.23){12}{\line(0,1){0.23}}
\multiput(38.56,32.72)(-0.11,0.24){11}{\line(0,1){0.24}}
\multiput(37.32,35.34)(-0.12,0.28){9}{\line(0,1){0.28}}
\multiput(36.27,37.84)(-0.11,0.30){8}{\line(0,1){0.30}}
\multiput(35.42,40.24)(-0.11,0.38){6}{\line(0,1){0.38}}
\multiput(34.77,42.52)(-0.11,0.54){4}{\line(0,1){0.54}}
\multiput(34.32,44.70)(-0.09,0.69){3}{\line(0,1){0.69}}
\put(34.06,46.76){\line(0,1){1.96}}
\multiput(34.00,48.72)(0.07,0.92){2}{\line(0,1){0.92}}
\multiput(34.14,50.56)(0.11,0.58){3}{\line(0,1){0.58}}
\multiput(34.48,52.30)(0.11,0.33){5}{\line(0,1){0.33}}
\multiput(35.01,53.93)(0.10,0.22){7}{\line(0,1){0.22}}
\multiput(35.74,55.44)(0.12,0.18){8}{\line(0,1){0.18}}
\multiput(36.67,56.85)(0.11,0.13){10}{\line(0,1){0.13}}
\multiput(37.79,58.15)(0.14,0.12){16}{\line(1,0){0.14}}
\put(22.67,25.00){\makebox(0,0)[cc]{observer}}
\put(55.00,30.00){\makebox(0,0)[cc]{object}}
\put(41.00,56.67){\makebox(0,0)[lc]{cartesian cut}}
\put(35.00,20.00){\vector(2,1){15.00}}
\put(46.67,34.00){\vector(-2,-1){14.67}}
\put(48.33,22.67){\makebox(0,0)[cc]{${\cal T}$}}
\put(31.33,32.33){\makebox(0,0)[cc]{${\cal S}$}}
\end{picture}
\end{center}
\caption{An interface as a  cartesian cut between observer and object.
The information flow across the interface is formalized by symbols.
\label{f1-cut}}
\end{figure}
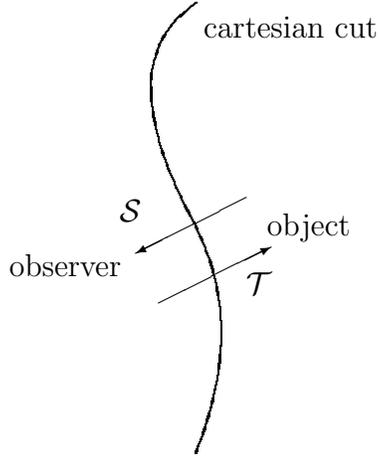

Informally, in a measurement, the object ``feels'' the observer's question (in ${\cal T}$)
and responds with an answer (in ${\cal S}$) which is felt by the observer
(cf. Fig. \ref{f1-cut}).

The reader is encouraged to view the interface not as a static
entity but as a {\em dynamic} one, through which information is constantly piped back and forth
the observer and the object and the resulting time flow may also be viewed as
the dynamic evolution of the system as a whole.
In what follows it is important to stress that
we shall restrict our attention to cases for which the
interface is total; i.e., the observer
receives {\em all} symbols emanating from the object.

\section*{One-to-one quantum state evolution and ``haunted'' measurements}

On a microphysical scale,
we do not wish to restrict quantum object symbols to
classical states. The concept pursued here is rather
that of the quantum scenario III: a uniform quantum system with unitary, and thus reversible, one-to-one
evolution.
Any process within the entire system evolves according to a reversible
law represented by a unitary time evolution $U^{-1}=U^\dagger$.
As a result, the interface map $I$ is one-to-one; i.e., it is a bijection.

Stated pointedly, we take it for granted that the wave
function of the entire system---including the observer and the
observed object separated by the cartesian cut or interface---evolves
one-to-one. Thus,  in principle, previous states can be reconstructed
by  proper reversible manipulations.

In this scenario, what is called ``measurement'' is merely an
exchange of quantum information.
In particular, the observer can ``undo'' a measurement by proper input
of quantum information via the quantum interface. In such a case, no
information, no knowledge about the object's state can remain on the
observer's side of the cut; all information has to be ``recycled''
completely in order
to be able to restore the wave function of the object entirely in its
previous form.

Experiments of the above form have been suggested
\cite{greenberger2}
and performed under the name ``haunted measurement''
and ``quantum eraser''  \cite{hkwz}.
These matters are very similar to the
opening,
closing and reopening of Schr\"odinger's catalogue of expectation values
\cite[p. 823]{schrodinger}:
At least up to
a certain magnitude of complexity, any measurement can be ``undone'' by
a proper reconstruction of the wave-function. A necessary condition for
this to happen is that {\em all} information about the original
measurement is lost.
In Schr\"odinger's terms,
the prediction catalog
(the wave function) can be opened only at one particular page.
We may close
the prediction catalog
before reading this page. Then we can open
the prediction catalog
at another, complementary, page again.
By no way we can open
the prediction catalog at one page, read and (irreversible) memorize the
page, close it; then open it at another, complementary, page.
(Two non-complementary pages which correspond to two co-measurable
observables can be read simultaneously.)

\section*{Where exactly is the interface located?}

The interface has been  introduced here as a scaffolding,
an auxiliary construction to model the information exchange
between the observer and the observed object.
One could quite justifyable ask
(and this question {\em has} indeed been asked by Professor Bryce deWitt),
``where exactly {\em is} the interface in a concrete experiment,
such as a spin state measurement in a Stern-Gerlach apparatus?''

We take the position here that the location of the interface very much depends
on the physical proposition which is tested and on the conventions assumed.
Let us take, for example, a statement  like
$${\textrm{``the electron spin in the $z$-direction is up.''}}$$

In the case of a Stern-Gerlach device, one could
locate the interface at the apparatus itself.
Then, the information passing through the interface is identified with
the way the particle took.

\begin{figure}
\begin{center}
\unitlength 1.00mm
\linethickness{0.4pt}
\begin{picture}(115.67,23.67)
\put(5.00,10.00){\circle{10.00}}
\put(11.67,10.00){\line(1,0){9.67}}
\put(21.33,10.00){\line(1,0){5.00}}
\multiput(26.33,10.00)(1.23,0.07){2}{\line(1,0){1.23}}
\multiput(28.80,10.14)(1.22,0.12){2}{\line(1,0){1.22}}
\multiput(31.24,10.38)(0.81,0.11){3}{\line(1,0){0.81}}
\multiput(33.65,10.71)(0.60,0.11){4}{\line(1,0){0.60}}
\multiput(36.05,11.14)(0.47,0.10){5}{\line(1,0){0.47}}
\multiput(38.42,11.67)(0.39,0.10){6}{\line(1,0){0.39}}
\multiput(40.76,12.28)(0.39,0.12){6}{\line(1,0){0.39}}
\multiput(43.08,13.00)(0.33,0.12){7}{\line(1,0){0.33}}
\multiput(45.38,13.81)(0.28,0.11){8}{\line(1,0){0.28}}
\multiput(47.65,14.71)(0.25,0.11){9}{\line(1,0){0.25}}
\multiput(49.90,15.71)(0.22,0.11){10}{\line(1,0){0.22}}
\multiput(52.13,16.81)(0.22,0.12){10}{\line(1,0){0.22}}
\put(21.33,10.00){\line(1,0){2.92}}
\put(24.25,9.97){\line(1,0){2.83}}
\multiput(27.08,9.86)(1.37,-0.08){2}{\line(1,0){1.37}}
\multiput(29.83,9.70)(1.33,-0.12){2}{\line(1,0){1.33}}
\multiput(32.49,9.46)(0.86,-0.10){3}{\line(1,0){0.86}}
\multiput(35.06,9.16)(0.62,-0.09){4}{\line(1,0){0.62}}
\multiput(37.55,8.78)(0.60,-0.11){4}{\line(1,0){0.60}}
\multiput(39.95,8.35)(0.46,-0.10){5}{\line(1,0){0.46}}
\multiput(42.26,7.84)(0.45,-0.11){5}{\line(1,0){0.45}}
\multiput(44.49,7.27)(0.36,-0.11){6}{\line(1,0){0.36}}
\multiput(46.62,6.62)(0.34,-0.12){6}{\line(1,0){0.34}}
\multiput(48.68,5.92)(0.28,-0.11){7}{\line(1,0){0.28}}
\multiput(50.64,5.14)(0.23,-0.11){8}{\line(1,0){0.23}}
\multiput(52.52,4.29)(0.23,-0.12){11}{\line(1,0){0.23}}
\put(56.00,22.00){\line(2,-3){2.67}}
\put(58.67,3.67){\line(-3,-5){2.40}}
\multiput(57.33,20.00)(0.40,0.11){6}{\line(1,0){0.40}}
\put(59.75,20.63){\line(1,0){2.14}}
\multiput(61.89,20.74)(0.47,-0.10){4}{\line(1,0){0.47}}
\multiput(63.75,20.34)(0.20,-0.12){8}{\line(1,0){0.20}}
\multiput(65.33,19.41)(0.12,-0.13){11}{\line(0,-1){0.13}}
\multiput(66.64,17.96)(0.11,-0.22){9}{\line(0,-1){0.22}}
\multiput(67.67,16.00)(0.12,-0.16){11}{\line(0,-1){0.16}}
\multiput(68.94,14.22)(0.14,-0.12){12}{\line(1,0){0.14}}
\multiput(70.61,12.82)(0.23,-0.11){9}{\line(1,0){0.23}}
\multiput(72.71,11.80)(0.52,-0.11){7}{\line(1,0){0.52}}
\multiput(57.67,1.33)(0.64,-0.11){4}{\line(1,0){0.64}}
\put(60.24,0.87){\line(1,0){2.23}}
\multiput(62.46,0.92)(0.38,0.11){5}{\line(1,0){0.38}}
\multiput(64.35,1.48)(0.17,0.12){9}{\line(1,0){0.17}}
\multiput(65.90,2.55)(0.12,0.19){15}{\line(0,1){0.19}}
\multiput(67.67,5.33)(0.12,0.26){6}{\line(0,1){0.26}}
\multiput(68.38,6.89)(0.14,0.12){9}{\line(1,0){0.14}}
\multiput(69.66,7.96)(0.37,0.12){5}{\line(1,0){0.37}}
\multiput(71.52,8.54)(2.40,-0.10){2}{\line(1,0){2.40}}
\put(76.67,7.00){\framebox(17.67,5.67)[cc]{}}
\put(78.33,12.67){\framebox(14.33,11.00)[cc]{UP}}
\put(19.67,15.00){\framebox(20.33,4.67)[cc]{}}
\put(19.67,0.33){\framebox(20.33,4.67)[cc]{}}
\put(105.00,15.00){\circle*{1.33}}
\put(110.00,15.00){\circle*{1.33}}
\put(115.00,15.00){\circle*{1.33}}
\put(56.00,22.00){\line(-3,-2){4.33}}
\put(58.67,18.00){\line(-3,-2){4.33}}
\put(56.33,-0.33){\line(-2,1){4.50}}
\put(58.67,3.67){\line(-5,3){4.83}}
\put(1.50,13.50){\line(1,-1){7.00}}
\put(8.50,13.50){\line(-1,-1){7.00}}
\end{picture}
\end{center}
\caption{Where exactly is the interface located?
\label{where-is-interface}}
\end{figure}
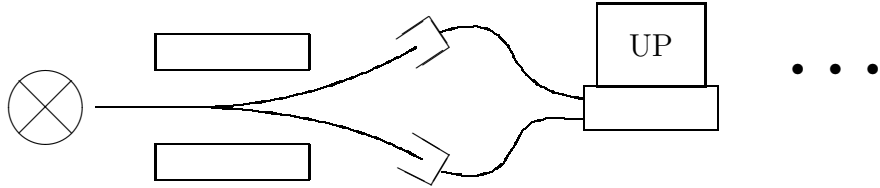

One could also locate the interface at two detectors at the end of the beam paths.
In this case, the informaton penetrating through the interface corresponds
to which one of the two detectors (assumed lossles) clicks (cf. Fig. \ref{where-is-interface}).

One could also situate the interface at the computer interface card registering this
click, or at an
experimenter who presumably monitors the event
(cf. Wigner's friend \cite{wigner:mb}), or at the persons of the research group
to whom the experimenter reports, to their scientific peers, and so on.

Since there is no material or real substrate which could be uniquely identified with
the interface, in principle it could be associated with or located at anything
which is affected
by the state of the object.
The only difference is the reconstructibility of the object's previous state (cf. below):
the ``more macroscopic'' (i.e., many-to-one) the interface becomes, the more difficult
it becomes to reconstruct the original state of the object.

\section*{From one-to-one to many-to-one}

If the quantum evolution is reversible,
how come that observers usually experience irreversibility in
measurement processes? We take the position here that {\em the concept
of irreversible measurement is no deep principle but merely originates
in the practical inability to reconstruct a quantum state of the object.}

Restriction to classical state or information exchange across the quantum
interface---the quasi-classical scenario II---effectively
implements the standard quantum description of the measurement process by
a classical measurement apparatus: there exists a clear distinction
between the ``internal quantum box,'' the quantum object---with unitary,
reversible, one-to-one internal evolution---and the classical symbols
emanating from it. Such a reduction from the quantum to the classical
world is accompanied by a loss of internal information
``carried with the quantum state.''
This effectively induces a many-to-one transition associated with the
measurement process, often referred to as ``wave function
collapse.''
In such a case, one and the same object symbol could have resulted from
many different quantum states, thereby giving raise to
irreversibility and entropy increase.

But also in the case of a uniform one-to-one evolution (scenario I),
just as in classical statistical physics, reconstruction greatly depends
on the possibility to ``keep track'' of all the information flow directed at and
emanating from the object. If  this flow is great and
spreads quickly with respect to the capabilities of the experimenter,
and if the reverse flow of information from the observer to the
object through the interface cannot be suitably controlled
\cite{zurek-91,zurek-98}
then the chances for reconstruction are low.

This is particularly true if the interface is not total:
in such a case, information flows off the object to regions which are
(maybe permanently) outside of the observer's control.

The possibility to reconstruct a particular state may widely vary
with technological capabilities which often boil down to financial
commitments.
Thus, irreversibility
of quantum measurements by interfaces appears as a gradual concept,
depending on conventions and practical necessities, and not
as a principal property of the quantum.

In terms of coding theory, the quantum object code is sent to the interface
but is not properly interpreted by the observer.
Indeed, the observer might only be able to understand a ``higher,''
macroscopic level of physical description, which subsumes several distinct
microstates under one macro-symbol (cf. below).
As a result, such macro-symbols are no
unique encoding of the object symbols.
Thus effectively the interface map $I$ becomes many-to-one.

This also elucidates the question
why there should be any meaningful concept of classical information if there is merely
quantum information to begin with: in such a scenario, classical information appears as
an effective entity on higher, intermediate levels of description.
Yet, the most fundamental level is quantum information.

\subsection*{Do conscious observers ``unthink''?}

Because of the one-to-one evolution,
a necessary condition for reconstruction of the object wave function is
the complete restoration of the observer wave function as well.
That is, the observer's state is restored to
its previous  form, and no knowledge, no trace whatsoever can be left
behind.
An observer would not even know that a ``measurement'' has taken place.
This is hard to accept, in particular if one assumes that observers
have consciousness which are detached entities from and not mere
functions of the quantum brain. Thus, in the latter case, one might be
convinced that conscious observers ``unthink'' the measurement results
in the process of complete restoration of the wave function. In the
latter case, consciousness might ``carry away'' the measurement result
via a process distinct from the quantum brain. (Cf. Wigner's friend
\cite{wigner:mb}.)

But even in this second,
dualistic, scenario, the conscious observer, after reconstruction of the
wave function, would have no direct proof of the ``previously measured fact,'' although
subsequent measurements might confirm his allegations.
This amounts to a proposal of an experiment involving
a conscious observer ({\em not} merely a rational agent) and
a quantized object.
The experiment tests the  metaphysical claim that consciousness exists beyond
matter \cite{eccles:papal}.
As sketched above, the experiment involves four steps.
\begin{itemize}
\item Step I: The conscious observer
measures some quantum observable on the quantized object which occurs irreducibly
random according to the axioms of quantum theory.
As a consequence, the observer ``is aware of''
the measurement result and ascribes to it an ``element of physical reality'' \cite{epr}.
\item Step II:
The original quantum state of the quantized object is reconstructed.
Thereby, all physical information about the measurement result is lost.
This is also true for the brain of the conscious observer.
Let us assume that the observer ``is still aware of''
the measurement result.
In this case,  the observer ascribes to it an ``element of metaphysical reality.''
\item Step III:
The observer guesses or predicts the outcome of the measurement despite the fact that
no empirical evidence about the outcome of the previous measurement exists.
\item Step IV:
The measurement is ``re-done'' and the actual measurement result is compared with the
conscious observer's prediction in step III.
If the prediction and the actual outcome do not coincide, the hypothesis
of a consciousness beyond matter is falsified.
\end{itemize}

As an analogy, one might think of a player in a virtual reality
environment. Although at the observation level of the virtual reality,
the measurement is undone, the player himself ``knows'' what has been
there before. This knowledge, however, has been passed on to
another interface
which is not immanent with respect to the virtual reality. That is, it
cannot be defined by intrinsic (endo-) means. Therefore, it can be
called a {\em transcendent interface} with respect to the virtual
reality. However, if we start with the real universe of the player, then
the same interface becomes intrinsically definable. The
hierarchical structure of meta-worlds has been the subject of
conceptual and visual art
\cite{weibel,totalrecall,Matrix} and literature \cite{simula}.

\subsection*{Parallels in statistical physics: from reversibility to
irreversibility}

The issue of ``emergence'' of irreversibility from reversible laws is an
old one and subject of scientific debate at least since Boltzmann's time
\cite{bricmont}.
We shall  shortly review an explanation in terms of the emergence
of many-to-one (irreversible) evolution relative to a ``higher''
macroscopic level of description from one-to-one
(reversible)
evolution at a more fundamental microscopic ``complete'' level of
description. These considerations are based on the work of
Jaynes \cite{jaynes2,jaynes-prob}, Katz \cite{katz} and  Hobson \cite{hobson}, among
others. See Bu\v{c}ek {\it et al.} \cite{buzek99} for a detailed review
with applications.

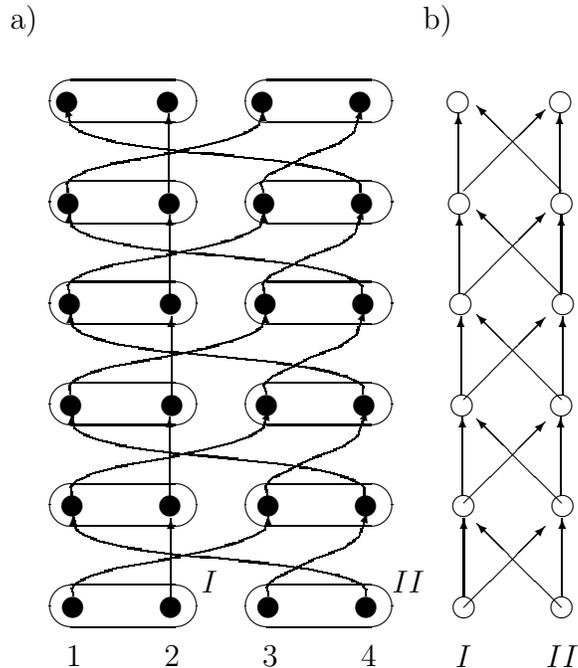
\begin{figure}
\begin{center}
\unitlength 1.30mm
\linethickness{0.4pt}
\begin{picture}(51.21,65.00)
\put(40.00,5.06){\circle{2.11}}
\put(50.00,5.06){\circle{2.11}}
\put(0.00,5.06){\circle*{2.11}}
\put(10.00,5.06){\circle*{2.11}}
\put(20.00,5.06){\circle*{2.11}}
\put(30.00,5.06){\circle*{2.11}}
\put(39.96,15.35){\circle{2.11}}
\put(50.10,15.35){\circle{2.11}}
\put(-0.04,15.35){\circle*{2.11}}
\put(9.96,15.35){\circle*{2.11}}
\put(19.96,15.35){\circle*{2.11}}
\put(29.96,15.35){\circle*{2.11}}
\put(40.00,6.11){\vector(0,1){8.02}}
\put(50.05,6.11){\vector(0,1){8.02}}
\put(40.00,0.01){\makebox(0,0)[cc]{$I$}}
\put(50.05,0.01){\makebox(0,0)[cc]{$II$}}
\put(0.11,0.01){\makebox(0,0)[cc]{$1$}}
\put(10.17,0.01){\makebox(0,0)[cc]{$2$}}
\put(20.22,0.01){\makebox(0,0)[cc]{$3$}}
\put(30.28,0.01){\makebox(0,0)[cc]{$4$}}
\put(29.96,6.15){\line(0,1){0.64}}
\multiput(30.02,6.79)(-0.11,0.12){5}{\line(0,1){0.12}}
\multiput(29.47,7.40)(-0.23,0.11){5}{\line(-1,0){0.23}}
\multiput(28.30,7.96)(-0.36,0.11){5}{\line(-1,0){0.36}}
\multiput(26.52,8.49)(-0.48,0.10){5}{\line(-1,0){0.48}}
\multiput(24.12,8.98)(-0.75,0.11){4}{\line(-1,0){0.75}}
\multiput(21.11,9.42)(-1.02,0.11){6}{\line(-1,0){1.02}}
\put(0.05,14.26){\vector(0,1){0.2}}
\multiput(15.01,10.06)(-0.99,0.09){4}{\line(-1,0){0.99}}
\multiput(11.04,10.42)(-0.84,0.10){4}{\line(-1,0){0.84}}
\multiput(7.68,10.84)(-0.69,0.12){4}{\line(-1,0){0.69}}
\multiput(4.93,11.32)(-0.43,0.11){5}{\line(-1,0){0.43}}
\multiput(2.78,11.85)(-0.31,0.12){5}{\line(-1,0){0.31}}
\multiput(1.23,12.45)(-0.16,0.11){6}{\line(-1,0){0.16}}
\multiput(0.29,13.10)(-0.08,0.39){3}{\line(0,1){0.39}}
\multiput(0.05,6.15)(-0.07,0.39){2}{\line(0,1){0.39}}
\multiput(-0.09,6.93)(0.12,0.12){6}{\line(0,1){0.12}}
\multiput(0.61,7.68)(0.25,0.12){6}{\line(1,0){0.25}}
\multiput(2.13,8.39)(0.39,0.11){6}{\line(1,0){0.39}}
\multiput(4.49,9.06)(0.61,0.11){9}{\line(1,0){0.61}}
\put(20.04,14.26){\vector(0,1){0.2}}
\multiput(9.98,10.06)(0.51,0.11){7}{\line(1,0){0.51}}
\multiput(13.58,10.81)(0.40,0.11){7}{\line(1,0){0.40}}
\multiput(16.40,11.56)(0.29,0.11){7}{\line(1,0){0.29}}
\multiput(18.43,12.31)(0.18,0.11){7}{\line(1,0){0.18}}
\multiput(19.68,13.06)(0.09,0.30){4}{\line(0,1){0.30}}
\multiput(20.04,6.15)(-0.08,0.29){3}{\line(0,1){0.29}}
\multiput(19.80,7.01)(0.11,0.11){9}{\line(0,1){0.11}}
\multiput(20.79,8.05)(0.24,0.12){17}{\line(1,0){0.24}}
\put(29.96,14.12){\vector(1,3){0.2}}
\multiput(24.93,10.06)(0.31,0.12){9}{\line(1,0){0.31}}
\multiput(27.70,11.12)(0.15,0.11){11}{\line(1,0){0.15}}
\multiput(29.38,12.37)(0.12,0.35){5}{\line(0,1){0.35}}
\put(10.04,6.11){\vector(0,1){8.02}}
\put(39.82,25.69){\circle{2.11}}
\put(39.68,36.03){\circle{2.11}}
\put(39.54,46.38){\circle{2.11}}
\put(39.40,56.72){\circle{2.11}}
\put(49.96,25.69){\circle{2.11}}
\put(50.15,36.03){\circle{2.11}}
\put(50.01,46.38){\circle{2.11}}
\put(49.87,56.72){\circle{2.11}}
\put(-0.18,25.69){\circle*{2.11}}
\put(-0.32,36.03){\circle*{2.11}}
\put(-0.46,46.38){\circle*{2.11}}
\put(-0.60,56.72){\circle*{2.11}}
\put(10.15,25.69){\circle*{2.11}}
\put(10.01,36.03){\circle*{2.11}}
\put(9.87,46.38){\circle*{2.11}}
\put(9.73,56.72){\circle*{2.11}}
\put(19.82,25.69){\circle*{2.11}}
\put(19.68,36.03){\circle*{2.11}}
\put(19.54,46.38){\circle*{2.11}}
\put(19.40,56.72){\circle*{2.11}}
\put(29.82,25.69){\circle*{2.11}}
\put(29.68,36.03){\circle*{2.11}}
\put(29.54,46.38){\circle*{2.11}}
\put(29.40,56.72){\circle*{2.11}}
\put(39.86,16.45){\vector(0,1){8.02}}
\put(39.72,26.79){\vector(0,1){8.02}}
\put(39.58,37.14){\vector(0,1){8.02}}
\put(39.44,47.48){\vector(0,1){8.02}}
\put(49.91,16.45){\vector(0,1){8.02}}
\put(50.10,26.79){\vector(0,1){8.02}}
\put(49.96,37.14){\vector(0,1){8.02}}
\put(49.82,47.48){\vector(0,1){8.02}}
\put(29.82,16.49){\line(0,1){0.64}}
\multiput(29.88,17.13)(-0.11,0.12){5}{\line(0,1){0.12}}
\multiput(29.33,17.74)(-0.23,0.11){5}{\line(-1,0){0.23}}
\multiput(28.16,18.30)(-0.36,0.11){5}{\line(-1,0){0.36}}
\multiput(26.38,18.83)(-0.48,0.10){5}{\line(-1,0){0.48}}
\multiput(23.98,19.32)(-0.75,0.11){4}{\line(-1,0){0.75}}
\multiput(20.97,19.77)(-1.02,0.11){6}{\line(-1,0){1.02}}
\put(29.68,26.84){\line(0,1){0.64}}
\multiput(29.74,27.48)(-0.11,0.12){5}{\line(0,1){0.12}}
\multiput(29.19,28.08)(-0.23,0.11){5}{\line(-1,0){0.23}}
\multiput(28.02,28.65)(-0.36,0.11){5}{\line(-1,0){0.36}}
\multiput(26.24,29.17)(-0.48,0.10){5}{\line(-1,0){0.48}}
\multiput(23.84,29.66)(-0.75,0.11){4}{\line(-1,0){0.75}}
\multiput(20.83,30.11)(-1.02,0.11){6}{\line(-1,0){1.02}}
\put(29.54,37.18){\line(0,1){0.64}}
\multiput(29.60,37.82)(-0.11,0.12){5}{\line(0,1){0.12}}
\multiput(29.05,38.43)(-0.23,0.11){5}{\line(-1,0){0.23}}
\multiput(27.88,38.99)(-0.36,0.11){5}{\line(-1,0){0.36}}
\multiput(26.10,39.52)(-0.48,0.10){5}{\line(-1,0){0.48}}
\multiput(23.70,40.01)(-0.75,0.11){4}{\line(-1,0){0.75}}
\multiput(20.69,40.45)(-1.02,0.11){6}{\line(-1,0){1.02}}
\put(29.40,47.52){\line(0,1){0.64}}
\multiput(29.46,48.16)(-0.11,0.12){5}{\line(0,1){0.12}}
\multiput(28.91,48.77)(-0.23,0.11){5}{\line(-1,0){0.23}}
\multiput(27.74,49.33)(-0.36,0.11){5}{\line(-1,0){0.36}}
\multiput(25.96,49.86)(-0.48,0.10){5}{\line(-1,0){0.48}}
\multiput(23.56,50.35)(-0.75,0.11){4}{\line(-1,0){0.75}}
\multiput(20.55,50.80)(-1.02,0.11){6}{\line(-1,0){1.02}}
\put(-0.09,24.60){\vector(0,1){0.2}}
\multiput(14.87,20.41)(-0.99,0.09){4}{\line(-1,0){0.99}}
\multiput(10.90,20.77)(-0.84,0.10){4}{\line(-1,0){0.84}}
\multiput(7.54,21.19)(-0.69,0.12){4}{\line(-1,0){0.69}}
\multiput(4.79,21.66)(-0.43,0.11){5}{\line(-1,0){0.43}}
\multiput(2.64,22.20)(-0.31,0.12){5}{\line(-1,0){0.31}}
\multiput(1.09,22.79)(-0.16,0.11){6}{\line(-1,0){0.16}}
\multiput(0.15,23.44)(-0.08,0.39){3}{\line(0,1){0.39}}
\put(-0.23,34.94){\vector(0,1){0.2}}
\multiput(14.73,30.75)(-0.99,0.09){4}{\line(-1,0){0.99}}
\multiput(10.76,31.11)(-0.84,0.10){4}{\line(-1,0){0.84}}
\multiput(7.40,31.53)(-0.69,0.12){4}{\line(-1,0){0.69}}
\multiput(4.65,32.00)(-0.43,0.11){5}{\line(-1,0){0.43}}
\multiput(2.50,32.54)(-0.31,0.12){5}{\line(-1,0){0.31}}
\multiput(0.95,33.13)(-0.16,0.11){6}{\line(-1,0){0.16}}
\multiput(0.01,33.78)(-0.08,0.39){3}{\line(0,1){0.39}}
\put(-0.37,45.29){\vector(0,1){0.2}}
\multiput(14.59,41.09)(-0.99,0.09){4}{\line(-1,0){0.99}}
\multiput(10.62,41.45)(-0.84,0.10){4}{\line(-1,0){0.84}}
\multiput(7.26,41.87)(-0.69,0.12){4}{\line(-1,0){0.69}}
\multiput(4.51,42.35)(-0.43,0.11){5}{\line(-1,0){0.43}}
\multiput(2.36,42.88)(-0.31,0.12){5}{\line(-1,0){0.31}}
\multiput(0.81,43.48)(-0.16,0.11){6}{\line(-1,0){0.16}}
\multiput(-0.13,44.13)(-0.08,0.39){3}{\line(0,1){0.39}}
\put(-0.51,55.63){\vector(-1,3){0.2}}
\multiput(14.45,51.44)(-1.32,0.12){3}{\line(-1,0){1.32}}
\multiput(10.48,51.80)(-0.84,0.10){4}{\line(-1,0){0.84}}
\multiput(7.12,52.21)(-0.69,0.12){4}{\line(-1,0){0.69}}
\multiput(4.37,52.69)(-0.43,0.11){5}{\line(-1,0){0.43}}
\multiput(2.22,53.22)(-0.31,0.12){5}{\line(-1,0){0.31}}
\multiput(0.67,53.82)(-0.16,0.11){6}{\line(-1,0){0.16}}
\multiput(-0.27,54.47)(-0.08,0.39){3}{\line(0,1){0.39}}
\multiput(-0.09,16.49)(-0.07,0.39){2}{\line(0,1){0.39}}
\multiput(-0.23,17.27)(0.12,0.12){6}{\line(0,1){0.12}}
\multiput(0.47,18.02)(0.25,0.12){6}{\line(1,0){0.25}}
\multiput(1.99,18.73)(0.39,0.11){6}{\line(1,0){0.39}}
\multiput(4.35,19.40)(0.61,0.11){9}{\line(1,0){0.61}}
\multiput(-0.23,26.84)(-0.07,0.39){2}{\line(0,1){0.39}}
\multiput(-0.37,27.62)(0.12,0.12){6}{\line(0,1){0.12}}
\multiput(0.33,28.36)(0.25,0.12){6}{\line(1,0){0.25}}
\multiput(1.85,29.07)(0.39,0.11){6}{\line(1,0){0.39}}
\multiput(4.21,29.75)(0.61,0.11){9}{\line(1,0){0.61}}
\multiput(-0.37,37.18)(-0.07,0.39){2}{\line(0,1){0.39}}
\multiput(-0.51,37.96)(0.12,0.12){6}{\line(0,1){0.12}}
\multiput(0.19,38.71)(0.25,0.12){6}{\line(1,0){0.25}}
\multiput(1.71,39.42)(0.39,0.11){6}{\line(1,0){0.39}}
\multiput(4.07,40.09)(0.61,0.11){9}{\line(1,0){0.61}}
\multiput(-0.51,47.52)(-0.07,0.39){2}{\line(0,1){0.39}}
\multiput(-0.65,48.30)(0.12,0.12){6}{\line(0,1){0.12}}
\multiput(0.05,49.05)(0.25,0.12){6}{\line(1,0){0.25}}
\multiput(1.57,49.76)(0.39,0.11){6}{\line(1,0){0.39}}
\multiput(3.93,50.43)(0.61,0.11){9}{\line(1,0){0.61}}
\put(19.90,24.60){\vector(1,3){0.2}}
\multiput(9.84,20.41)(0.51,0.11){7}{\line(1,0){0.51}}
\multiput(13.44,21.16)(0.40,0.11){7}{\line(1,0){0.40}}
\multiput(16.26,21.90)(0.29,0.11){7}{\line(1,0){0.29}}
\multiput(18.29,22.65)(0.18,0.11){7}{\line(1,0){0.18}}
\multiput(19.54,23.40)(0.09,0.30){4}{\line(0,1){0.30}}
\put(19.76,34.94){\vector(0,1){0.2}}
\multiput(9.70,30.75)(0.51,0.11){7}{\line(1,0){0.51}}
\multiput(13.30,31.50)(0.40,0.11){7}{\line(1,0){0.40}}
\multiput(16.12,32.25)(0.29,0.11){7}{\line(1,0){0.29}}
\multiput(18.15,33.00)(0.18,0.11){7}{\line(1,0){0.18}}
\multiput(19.40,33.74)(0.09,0.30){4}{\line(0,1){0.30}}
\put(19.62,45.29){\vector(0,1){0.2}}
\multiput(9.56,41.09)(0.51,0.11){7}{\line(1,0){0.51}}
\multiput(13.16,41.84)(0.40,0.11){7}{\line(1,0){0.40}}
\multiput(15.98,42.59)(0.29,0.11){7}{\line(1,0){0.29}}
\multiput(18.01,43.34)(0.18,0.11){7}{\line(1,0){0.18}}
\multiput(19.26,44.09)(0.09,0.30){4}{\line(0,1){0.30}}
\put(19.48,55.63){\vector(0,1){0.2}}
\multiput(9.42,51.44)(0.51,0.11){7}{\line(1,0){0.51}}
\multiput(13.02,52.19)(0.40,0.11){7}{\line(1,0){0.40}}
\multiput(15.84,52.93)(0.29,0.11){7}{\line(1,0){0.29}}
\multiput(17.87,53.68)(0.18,0.11){7}{\line(1,0){0.18}}
\multiput(19.12,54.43)(0.09,0.30){4}{\line(0,1){0.30}}
\multiput(19.90,16.49)(-0.08,0.29){3}{\line(0,1){0.29}}
\multiput(19.66,17.35)(0.11,0.12){9}{\line(0,1){0.12}}
\multiput(20.65,18.39)(0.24,0.12){17}{\line(1,0){0.24}}
\multiput(19.76,26.84)(-0.08,0.29){3}{\line(0,1){0.29}}
\multiput(19.52,27.70)(0.11,0.11){9}{\line(0,1){0.11}}
\multiput(20.51,28.73)(0.24,0.12){17}{\line(1,0){0.24}}
\multiput(19.62,37.18)(-0.08,0.29){3}{\line(0,1){0.29}}
\multiput(19.38,38.04)(0.11,0.11){9}{\line(0,1){0.11}}
\multiput(20.37,39.08)(0.24,0.12){17}{\line(1,0){0.24}}
\multiput(19.48,47.52)(-0.08,0.29){3}{\line(0,1){0.29}}
\multiput(19.24,48.38)(0.11,0.12){9}{\line(0,1){0.12}}
\multiput(20.23,49.42)(0.24,0.12){17}{\line(1,0){0.24}}
\put(29.82,24.46){\vector(1,4){0.2}}
\multiput(24.79,20.41)(0.31,0.12){9}{\line(1,0){0.31}}
\multiput(27.56,21.46)(0.15,0.11){11}{\line(1,0){0.15}}
\multiput(29.24,22.71)(0.12,0.35){5}{\line(0,1){0.35}}
\put(29.68,34.80){\vector(1,3){0.2}}
\multiput(24.65,30.75)(0.31,0.12){9}{\line(1,0){0.31}}
\multiput(27.42,31.81)(0.15,0.11){11}{\line(1,0){0.15}}
\multiput(29.10,33.05)(0.12,0.35){5}{\line(0,1){0.35}}
\put(29.54,45.15){\vector(1,3){0.2}}
\multiput(24.51,41.09)(0.31,0.12){9}{\line(1,0){0.31}}
\multiput(27.28,42.15)(0.15,0.11){11}{\line(1,0){0.15}}
\multiput(28.96,43.40)(0.12,0.35){5}{\line(0,1){0.35}}
\put(29.40,55.49){\vector(1,3){0.2}}
\multiput(24.37,51.44)(0.31,0.12){9}{\line(1,0){0.31}}
\multiput(27.14,52.49)(0.15,0.11){11}{\line(1,0){0.15}}
\multiput(28.82,53.74)(0.12,0.35){5}{\line(0,1){0.35}}
\put(9.90,16.45){\vector(0,1){8.02}}
\put(10.09,26.79){\vector(0,1){8.02}}
\put(9.95,37.14){\vector(0,1){8.02}}
\put(9.81,47.48){\vector(0,1){8.02}}
\put(40.00,5.00){\vector(1,1){8.33}}
\put(50.00,5.00){\vector(-1,1){8.67}}
\put(40.00,15.67){\vector(1,1){8.33}}
\put(40.00,26.33){\vector(1,1){8.33}}
\put(40.00,37.00){\vector(1,1){8.33}}
\put(40.00,47.67){\vector(1,1){8.33}}
\put(50.00,15.67){\vector(-1,1){8.67}}
\put(50.00,26.33){\vector(-1,1){8.67}}
\put(50.00,37.00){\vector(-1,1){8.67}}
\put(50.00,47.67){\vector(-1,1){8.67}}
\put(5.17,5.17){\oval(15.00,4.33)[]}
\put(25.17,5.17){\oval(15.00,4.33)[]}
\put(5.17,15.50){\oval(15.00,4.33)[]}
\put(5.17,25.83){\oval(15.00,4.33)[]}
\put(5.17,36.17){\oval(15.00,4.33)[]}
\put(5.17,46.50){\oval(15.00,4.33)[]}
\put(5.17,56.83){\oval(15.00,4.33)[]}
\put(25.17,15.50){\oval(15.00,4.33)[]}
\put(25.17,25.83){\oval(15.00,4.33)[]}
\put(25.17,36.17){\oval(15.00,4.33)[]}
\put(25.17,46.50){\oval(15.00,4.33)[]}
\put(25.17,56.83){\oval(15.00,4.33)[]}
\put(-5.00,65.00){\makebox(0,0)[cc]{a)}}
\put(37.33,65.00){\makebox(0,0)[cc]{b)}}
\put(14.00,7.67){\makebox(0,0)[cc]{$I$}}
\put(34.33,7.67){\makebox(0,0)[cc]{$II$}}
\end{picture}
\end{center}
\caption{Full circles represents the ``complete'' level
of description, open circles or the corresponding ovals
represent intermediate or macroscopic levels of description.
The microphysical states $1,2$ are mapped onto $I$ and
$3,4$ are mapped onto $II$ by a many-to-one mapping.
a) The temporal evolution in terms of   the microstates is one-to-one;
b) The evolution with respect to the macro-states is irreversible.
\label{f2-mvm}}
\end{figure}

In this framework, the many-to-one and thus irreversible evolution
is a simple consequence of the fact that many different
microstates, i.e., states
on the fundamental ``complete'' level of physical description, are
mapped onto a {\em single} macroscopic state (cf. Fig. \ref{f2-mvm}).
Thereby, knowledge about the
microphysical state is lost; making impossible the later reconstruction
of the microphysical state from the macroscopic one.
(In the example drawn in Fig. \ref{f2-mvm}, observation of the ``macrostate''
$II$ could mean that the system is either in microstate $1$ or $2$.)
on some intermediate,
``higher'' level of physical description, whereas it remains reversible on
the complete description level.

Here, just as in the quantum interface case,
irreversibity in statistical physics is a gradual concept, very much
depending on the observation level, which depends on conventions and practical necessities.
Yet again, in principle the
underlying complete  level of description is one-to-one.
As a consequence, this would for example  make possible the
reconstruction of the Library of Alexandria if one takes into account
all smoky emanations thereof.
The task of ``reversing the gear,'' of reconstructing the past
and constructing a different future, is thus not entirely absurd.
Yet fortunately or unfortunately,
for all practical purposes it remains impossible.

\section*{Principle of information conservation}

In another scenario (closely related to scenario I),
classical information is a primary entity.
The quantum is obtained as an effective theory to represent the state of knowledge,
the ``knowables,''
of the observer about the object \cite{zeil-99,zeil-bruk-99,zeil-bruk-99a}.
Thereby, quantum information appears as a derived theoretical entity,
very much in the spirit of Schr\"odinger's perception of the wave function as a
catalogue of expectation values (cf. above).

The following circular definitions are assumed.
\begin{itemize}
\item
An elementary object carries one bit of (classical) information \cite{zeil-99}.
\item
$n$ elementary objects carry $n$ bits of (classical) information \cite{zeil-99}. The information content
present in the physical system {\em is exhausted} by the $n$ bits given;
nothing more can be gained by any perceivable procedure.
\item
Throughout temporal evolution, the amount of (classical) information measured in bits is conserved.
\end{itemize}

One immediate consequence seems a certain kind of irreducible randomness associated with
requesting from an elementary object information which has not been
previously encoded therein.
We may, for instance, think of an elementary object as an electron which has been prepared in spin state
``up'' in some direction. If the electron's spin state is measured in another direction, this
must give rise to randomness since the particle ``is not supposed to know'' about this property.
Yet, we may argue that in such a case the particle might respond with no answer at all, and not with the type of
irreducible randomness which, as we know from the computer sciences \cite{chaitin-99,calude:92b},
is such a preciously expensive quality.

One way to avoid this problem is to assume that the apparent randomness does not originate from the object but is a
property of the interface: the object always responds to the question it has been prepared for to answer; but the interface ``translates''
the observer's question into the appropriate form suitable for the object.
In this process, indeterminism comes in.

As a result of the assumption of the temporal conservation of information,
the evolution of the system has to be one-to-one and, for finite systems,
a permutation.

Another consequence of the conservation of information is the possibility to define continuity equations.
In analogy to magnetostatics or thermodynamics we may
represent the information flow by a vector which gives
the amount of information passing per unit area and per unit time through a surface element
at right angles to the flow.  We call this
the {\em information flow density} ${\bf j}$.
The amount
of information flowing across a small area $\Delta A$ in a unit time is
$${\bf j}\cdot {\bf n}\; \Delta A,$$
where ${\bf n}$ is the unit vector normal to $\Delta A$.
The information flow density is related to the average flow velocity $v$ of information.
In particular, the information flow density associated with
an elementary object of velocity $v$ per unit time is given by
${\bf j}= \rho v$ bits per second, where $\rho $ stands for the
information density (measured in bits/$m^3$).
For $N$ elementary objects per unit volume carrying one bit each,
$${\bf j}= Nvi.$$
Here, $i$ denotes the elementary quantity of information measured in bit units.
The information flow $I$ is the total amount of information passing per unit time
through any surface $A$; i.e.,
$$I=\int_A {\bf j}\cdot {\bf n} \;dA.$$

We have assumed that the cut is on a closed surface ${\cal A}_c$ surrounding the object.
The conservation law of information requires the following continuity equation to
be valid:
$$\int_{{\cal A}_c}{\bf j}\cdot {\bf n}\; dA = -{d\over dt}({\rm Information\; inside})$$
or, by defining an information density $\rho$ and applying Gauss' law,
$$\nabla \cdot {\bf j}=   -{d\rho \over dt}.$$

To give a quantitative account of the present
ability to reconstruct the quantum wave function of single photons,
we analyze the ``quantum eraser'' paper by Herzog, Kwiat, Weinfurter and Zeilinger
\cite{hkwz}. The authors report an extension of their apparatus of $x= 0.13$~m,
which amounts to an information passing through a sphere of radius $x$ of
$$I_{\rm qe}= {4 \pi x^2 ci}=6\times 10^7 {\rm bits/second}.$$
Here, ${\bf j}=ci$ ($c$ stands for the velocity of light in vacuum) has been assumed.
At this rate the reconstruction of the photon wave function has been conceivable.

We propose to consider $I$ as a measure for wave function reconstruction.
In general, $I$ will be astronomically high because of the astronomical numbers of
elementary objects involved. Yet, the associated diffusion velocity $v$ may be considerably lower than $c$.

Let us finally come back to the question,
{\em ``why should there be any meaningful concept of classical information if there is merely
quantum information to begin with?''}
A tentative  answer in the spirit of this approach
would be that {\em ``quantum information is merely a
concept derived from the necessity to formalize modes of thinking about the state of knowledge
of a classical observer about a classical object.
Although the interface is purely classical, it {\em appears} to the observer as if it were
purely quantum or quasi-classical.''}

\section*{Virtual reality as a quantum double}

Just as quantum systems, virtual reality universes can have a one-to-one
evolution. We shall shortly review reversible automata
\cite{sv-aut-rev,svo-oto} which are characterized by the following properties:
\begin{itemize}
\item
a finite set
$S$
of states,
\item
a finite input  alphabet $I$,
\item
a finite output alphabet $O$,
\item
temporal evolution function
$\delta :S\times I\rightarrow S$,
\item
output function
$\lambda :S\times I\rightarrow O$.
\end{itemize}
The combined transition and output function $U$ is reversible
and thus corresponds to a permutation:
\begin{equation}
U:(s,i)\rightarrow (\delta(s,i),\lambda (s,i)),
\label{t-e-l}
\end{equation}
with
$s\in S$ and $i\in I$.
Note that neither $\delta$ nor $\lambda$ needs to be a bijection.

As an example, take the perturbation matrix
$$
{U}=
\left(
\begin{array}{cccccc}
1&0&0&0&0&0 \\
0&1&0&0&0&0 \\
0&0&0&0&1&0 \\
0&0&0&1&0&0 \\
0&0&0&0&0&1 \\
0&0&1&0&0&0
\end{array}
\right).
$$
It can be realized by a reversible automaton which is represented in
Table
\ref{t-ra22}.
\begin{table}
\begin{center}
\begin{tabular}{|c|ccc|ccc|}
 \hline\hline
 &$\delta$ & && $\lambda$&&\\
$S\backslash I$ &1&2&3& 1&2&3\\
 \hline
$s_1$&$s_1$&$s_1$ &$s_2$ & 1&2&2\\
$s_2$&$s_2$&$s_2$ &$s_1$ & 1&3&3\\
 \hline\hline
\end{tabular}
\end{center}
\caption{Transition and output table of a reversible
automaton  with two states $S=\{s_1, s_2\}$ and three
input/output symbols $I= \{1,2,3\}$.
Neither its  transition nor its output function is one-to-one.
\label{t-ra22}}
\end{table}
Neither its
evolution function nor its transition function is one-to-one, since for
example
$
\delta (s_1,3)
=
\delta (s_2,1)=s_2
$
and
$
\lambda (s_1,2)
=
\lambda (s_1,3)=2
$.
Its flow diagram throughout five evolution steps is depicted in Figure
\ref{f2-mvm}, where the microstates $1,2,3,4$ are identified by
$(s_1,1)$,
$(s_1,2)$,
$(s_2,1)$ and
$(s_2,2)$, respectively.

\section*{Metaphysical speculations}

Although the contemporaries always attempt to canonize their
relative status of knowledge about the physical world, from a broader
historical perspective this appears sentimental at best and ridiculous
at worst.
The type of natural sciences which
emerged from the Enlightenment is in a permanent scientific revolution.
As a result, scientific wisdom is always transitory.
Science is and needs to be in constant change.

So, what about the quantum?
Quantum mechanics challenges the conventional rational understanding in
the following ways:
\begin{itemize}
\item
by allowing for randomness of single events, which collectively obey
quantum statistical predictions;
\item
by the feature of complementarity; i.e., the mutual exclusiveness
of the measurement of certain observables termed complementary.
Complementarity results in a non-classical, non-distributive and thus
non-boolean event structures;
\item
by non-standard probabilities which are based on non-classical,
non-boolean event structures. These quantum probabilities cannot be
properly composed from its proper parts, giving rise to the so-called
``contextuality.''
\end{itemize}

I believe that, just as so many other formalisms before, also
quantum theory
will eventually give way to a more comprehensive understanding of
fundamental
physics, although at the moment it appears almost heretic to pretend
that there is something ``beyond the quantum''.
Exactly how this progressive theory beyond the quantum will look like,
nobody
presently can say \cite{lakatosch}. (Otherwise, it would not be beyond
anymore, but there would be another theory lurking beyond the beyond.)
In view of the quantum challenges outlined before, it may be well
worth speculating that the revolution will  drastically change
our perception of the world.

It may well be that epistemic
issues such as the ones reviewed here will play an important role therein.
I believe that the careful analysis of conventions which are
taken for granted and are never mentioned in standard presentations of the
quantum and relativity theory \cite{svozil-relrel} will clarify some misconceptions.

Are quantum-like and relativity-like
theories  consequences of the modes we use to think about and construct
our world? Do they not tell us more about our projections than about an elusive reality?

Of course, physical constants such as Planck's constant
or the velocity of light {\em are} physical input.
But the {\em structural form} of the theories might be conventional.

Let me also state that one-to-one evolution is a sort of ``Borgesian''
 nightmare, a hermetic prison: the time
evolution is a constant permutation of one and the same ``message''
which always remains the same but expresses itself through different forms.
Information is neither created nor discarded but remains constant at all times.
The implicit time symmetry spoils the very notion of
``progress'' or ``achievement,''  since what
is a valuable output is purely determined by the
subjective meaning the observer associates with it and is devoid of any
syntactic relevance. In such a
scenario, any gain in knowledge remains a merely subjective
impression of ignorant observers.





\end{document}